\documentclass[prl, showpacs, superscriptaddress,
twocolumn,floatfix ]{revtex4}

\usepackage[OT1]{fontenc}
\usepackage[latin2]{inputenc}
\usepackage{graphicx,wrapfig}
\usepackage{color}
\newcommand{\BM}[1]{\mbox{\boldmath$#1$}}

\begin{document}

\title{Dislocation Glasses: Ageing during Relaxation and
Coarsening}

\author{B. Bak\'o}
\affiliation{E\"otv\"os University, Department of Materials Physics,
P\'azm\'any s\'et\'any 1/A, 1117 Budapest, Hungary}
\affiliation{Present address: Paul Scherrer Institute, 5232 Villigen PSI,
Switzerland}
%\email{botond.bako@psi.ch}

\author{I. Groma}
\affiliation{E\"otv\"os University, Department of Materials  Physics,
P\'azm\'any s\'et\'any 1/A, 1117 Budapest, Hungary}
%\email{groma@metal.elte.hu}

\author{G. Gy\"orgyi}
\affiliation{Institute for Theoretical Physics, HAS Research Groups, E\"otv\"os
University}
\email{gyorgyi@glu.elte.hu}

\author{G.T. Zim\'anyi}
\affiliation{Department of Physics, University of California, Davis}
%\email{zimanyi@physics.ucdavis.edu}

\begin{abstract} 
\noindent 
The dynamics of dislocations is reported to exhibit a range of glassy
properties. We study numerically various versions of 2d edge
dislocation systems, in the absence of externally applied stress.  Two
types of glassy behavior are identified (i) dislocations gliding along
randomly placed, but fixed, axes exhibit relaxation to their spatially
disordered stable state; (ii) if both climb and annihilation is
allowed, irregular cellular structures can form on a growing length
scale before all dislocations annihilate.  In all cases both the
correlation function and the diffusion coefficient are found to
exhibit ageing.  Relaxation in case (i) is a slow power law,
furthermore, in the transient process (ii) the dynamical exponent
$z\approx 6$, i.e., the cellular structure coarsens relatively slowly.

\end{abstract}

\maketitle

Since dislocations are ubiquitous in crystals and they can have a
dramatic effect on physical properties of materials, understanding the
morphology and dynamics of dislocations is of fundamental importance.
They are the carriers of plastic deformation in solids
\cite{kroener81} and are unwelcome impurities in single-crystal growth
\cite{rudolph05}.  In 2d they play a crucial role in melting
\cite{halperin78}, appear in Wigner crystals \cite{fisher79}, and
dislocations in vortex lattices \cite{blatter94} determine properties
of superconducting films \cite{miguel01}.  Further examples where
dislocations arise include magnetic bubble structures
\cite{seshadri92}, charge density waves \cite{gruner88}, colloidal
lattices \cite{murray87}, and dusty plasma \cite{quinn01}.

The morphology of dislocations is one of the most studied and least
understood aspects of crystal defects.  While there is a multitude of
dislocation patterns, one can discern a few generic types.  Such are
fractals and ladders in the presence of external shear
\cite{hahner98}, as well as cellular and diffuse configurations in
unsheared crystals.  The latter ones were discussed in the remarkable
recent experimental papers by Rudolph et al.\
\cite{rudolph05,rudolphetal05} with the conclusion that climb mobility
of dislocations has an important effect on morphology. Specifically,
when there is climb then cells of some characteristic length scale can
form, or, if climb is suppressed, dislocations freeze into a
diffuse-looking random distribution.  A recent work by the present
authors corroborated this finding on the basis of mesoscale
simulations \cite{bako06}.

Just a glance at disordered dislocation configurations from
experiments \cite{rudolph05} and simulations \cite{bako06,bulatov06}
suffices to raise the question, to what extent dislocations exhibit
glassy properties. Indications of slow equilibration came recently
from numerical studies of relaxing 2d dislocation systems
\cite{csikor05}.  A physical argument for the glass analogy is that
the long-range interaction between dislocations can change sign as
function of relative angles, thus making frustration possible.
Frustrated interactions are known to give rise to spin glasses in the
presence of quenched disorder \cite{young04}.  Furthermore, there is
an external element of randomness in dislocation systems that arises
from initial conditions. This reinforces the analogy, because random
initial conditions due to quenching from the high-$T$ state are an
essential ingredient in structural-glass-formers, like in
Lennard-Jones systems \cite{kob97}, and lead to slow growth of
irregular domains even in systems as simple as the Ising model
\cite{biroli05}.

There is a growing body of knowledge about the dynamics of glasses
(see \cite{biroli05} and refs.\ therein).  A central feature of glassy
dynamics is the dramatic slowing down while the temperature is lowered
\cite{tarjus05}.  Not only does the dynamics slow down, but it also
exhibits ageing, widely considered as a hallmark of glassy dynamics
\cite{kurchan97}. There are several qualitatively different types of
systems exhibiting ageing, such as (i) spin glasses with quenched
disordered interactions, (ii) systems approaching a non-glassy
equilibrium state, while random initial condition, possibly augmented
with kinetic constraints, slow down relaxation, and (iii) structural
glasses wherein long-time disorder originates in initial condition.

In this Letter we study the dynamics of edge dislocations in 2d by
mesoscale simulations \cite{bako06}, wherein dislocations are the
particles interacting by their stress fields and obeying overdamped
dynamics. While this is a strong simplification of real 3d dislocation
networks, for FCC crystals under certain conditions it is a reasonable
approximation, and it is of course relevant for defects in 2d
lattices.  Two main situations are considered, in the first one
dislocations only glide along the axis of their Burgers vector, in the
second one they can also climb perpendicular to it, and here we also
allow annihilation.  Climb has a dramatic effect, without it
dislocations relax to a (meta)stable state, while climb with
annihilation results in an increasingly dilute system and eventually
all dislocations vanish, corresponding to the dipolized state
predicted long time ago \cite{halperin78}.  In both situations we
study the cases of a single and three glide axes.  The effect of
temperature on ageing is tested by the addition of a Langevin force. 
 
Our main observations are as follows. (1) Dislocation configurations
after a fast initial transient depend on whether climb is possible and
on the number of glide axes.  With climb, walls form for both the 1-
and 3-slip systems, in the latter leading to a cellular structure with
a pronounced length scale, whereas without climb typical
configurations appear more disordered.  (2) All systems exhibit
ageing, that is, both the correlation function and the effective
diffusion coefficient of dislocations depend on the waiting time.  (3)
In the absence of climb the correlation functions decay by a power law
while a stable state is approached.  (4) Sufficiently high temperature
restores non-ageing dynamics. (5) With climb, the time exponent $z$ is
identified, characterizing the growth of the mean domain diameter. 
 
First we discuss the details of the simulation.  Three different
possible Burgers vectors defined by the directions 
$\pm(\cos(m\pi/3),\sin(m\pi/3))$, $m = 0,1,2$ in a square area of size
$L\times L$ was considered with periodic boundary conditions.  This
geometry emulates a hexagonal underlying lattice.  In order to
calculate the stress field of a dislocation with Burgers vector
$(b_x,b_y)$, we use the formalism developed by Kr\"oner
\cite{kroener81}.  The stress function $\chi (r)$ for this problem
fulfills the biharmonic equation 
\begin{eqnarray}
\nabla ^ 4 \chi = \mathcal{C}(b_x\partial_y -
b_y\partial_x)\delta(x)\delta(y),
\label{eq:biharmonic}\end{eqnarray} 
with $\mathcal{C}=2\mu/(1-\nu)$ in 3d and $\mathcal{C}=2\mu(1+\nu)$ in
2d, where $\mu$ is the shear modulus and $\nu$ Poisson's ratio. From
$\chi(r)$ the stress components are $\sigma_{xx} = -\partial_y
\partial_y\chi,\, \sigma_{yy} = -\partial_x \partial_x\chi, \,
\sigma_{xy} = \partial_x \partial_y\chi$.  The force $\BM F$ per unit
length acting on a dislocation with Burgers vector $\BM b$ is given by
the Peach-Koehler formula $\BM F = ({ \hat \sigma} \hspace{1mm}{ \BM
  b}) \times { \BM l}$ in which ${\BM l}=(0,0,1)$ and the stress $\hat
\sigma$ is due to all other dislocations. Equation
(\ref{eq:biharmonic}), discretized to $M^2$ points, with periodic
boundary conditions, can be solved for $\chi$ by using the Fourier
transform (FT) of the partial difference operator,
$D_m=\frac{2iM}{L}\sin\left(\frac{\pi m}{M}\right )$, where $m
=0,\dots,M-1$.  Then one obtains the FT of the stress function as
\begin{eqnarray}
\tilde\chi_{lm}=\mathcal{C}(b_x D_m- b_y D_l)/[D_l^2 + D_m^2]^2,
\end{eqnarray}
yielding the FT of the stress components 
$\tilde\sigma_{xx}= -D_l D_l \tilde\chi_{lm}$, 
$\tilde\sigma_{yy}= - D_m D_m \tilde\chi_{lm}$, 
$\tilde\sigma_{xy}= D_l D_m \tilde\chi_{lm}$. 
Inverse  FT gives the stress components at the grid points for
a single dislocation. For the effect of all dislocations, in $\tilde\chi_{lm}$
we replace  ${\BM b}$ by the FT of the discretized Burgers vector field.
While we keep track of the dislocation positions to high precision, the stress
field is discretized to $M=256$.  Note that this recipe suppresses interaction
between dislocations in the same box, but in our case such concurrence is rare.
By applying fast FT, for up to  10000 dislocations the above
algorithm is about 10 times faster than the direct calculation of the pair
interactions, for more details see \cite{bako06}.
Assuming overdamped dislocation dynamics, the ${\BM v}^i_g$ glide and ${\BM
v}^i_c$ climb velocity components of the $i$th dislocation are
\begin{eqnarray}
{\BM  v}^i_{g,c}  = \left[\Gamma_{g,c} ({\BM F}^i{\BM
a}^i_{g,c})+\sqrt{2T\Gamma_{g,c}}\xi_{g,c}\right]{\BM a}^i_{g,c},
\label{eq:v}
\end{eqnarray}
where ${\BM F}^i$ is the Peach-Koehler force given above, $\Gamma_{g}$
and $\Gamma_{c}$ are the glide and the climb mobilities, respectively,
${\BM a}_g={\BM b}_i/|{\BM b}|_i$, and ${\BM a}^i_c$ is a unit vector
perpendicular to ${\BM a}^i_g$.  Thermal fluctuations at temperature
$T$ are included as white noise $\xi_{g,c}$ with $\langle
\xi_{g,c}(t)\xi_{g,c}(t')\rangle=\delta(t-t')$. (So as to eliminate
material parameters the dimensionless variables ${\BM r} \Rightarrow
{\BM r}/L$, $t\Rightarrow t\Gamma_g\mathcal{C}b^2/4\pi L^2$, and
$T\Rightarrow 4\pi T/\mathcal{C}b^2$ are used henceforth.)
 
\begin{figure}
\begin{center}
\includegraphics[width=0.2\textwidth]{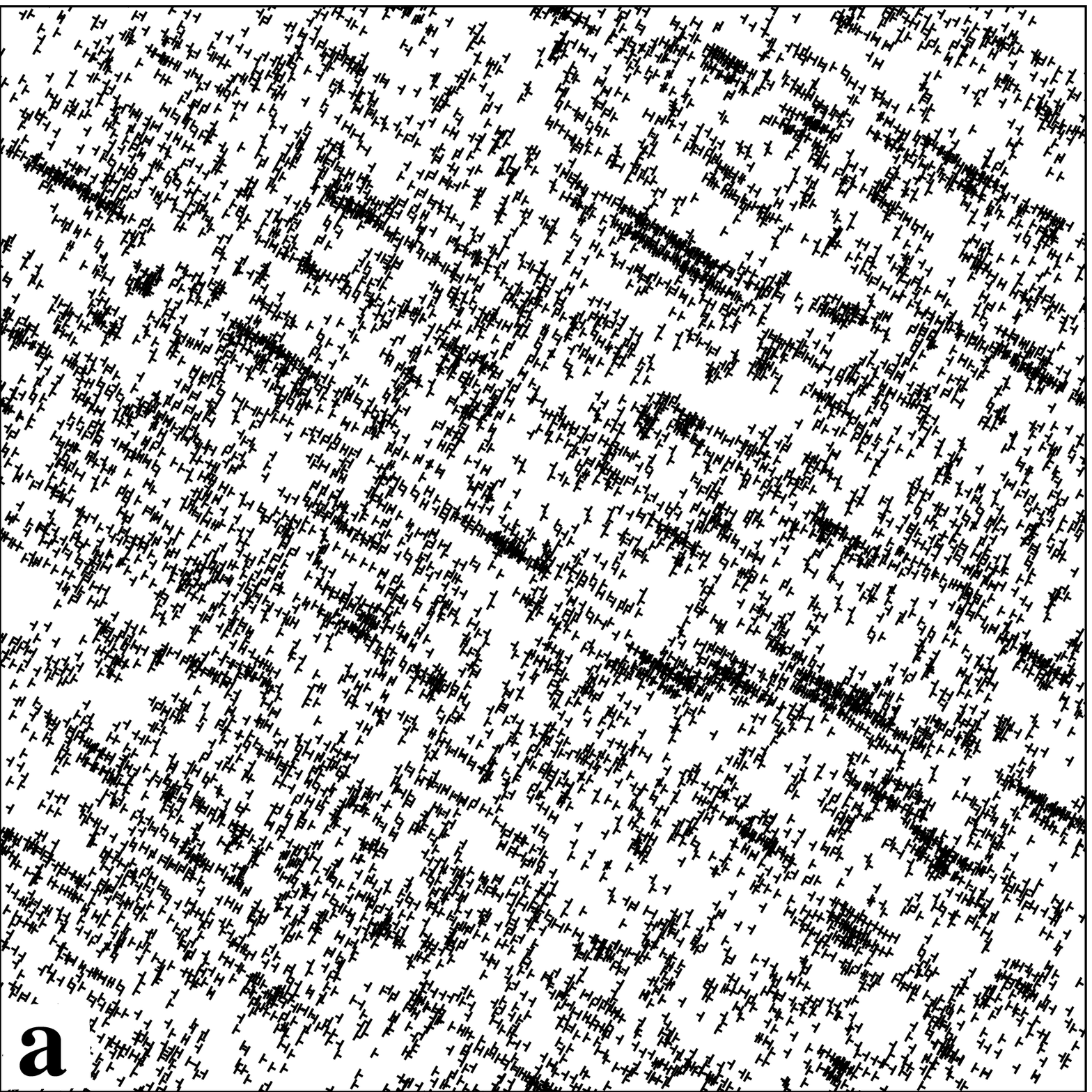}
\includegraphics[width=0.2\textwidth]{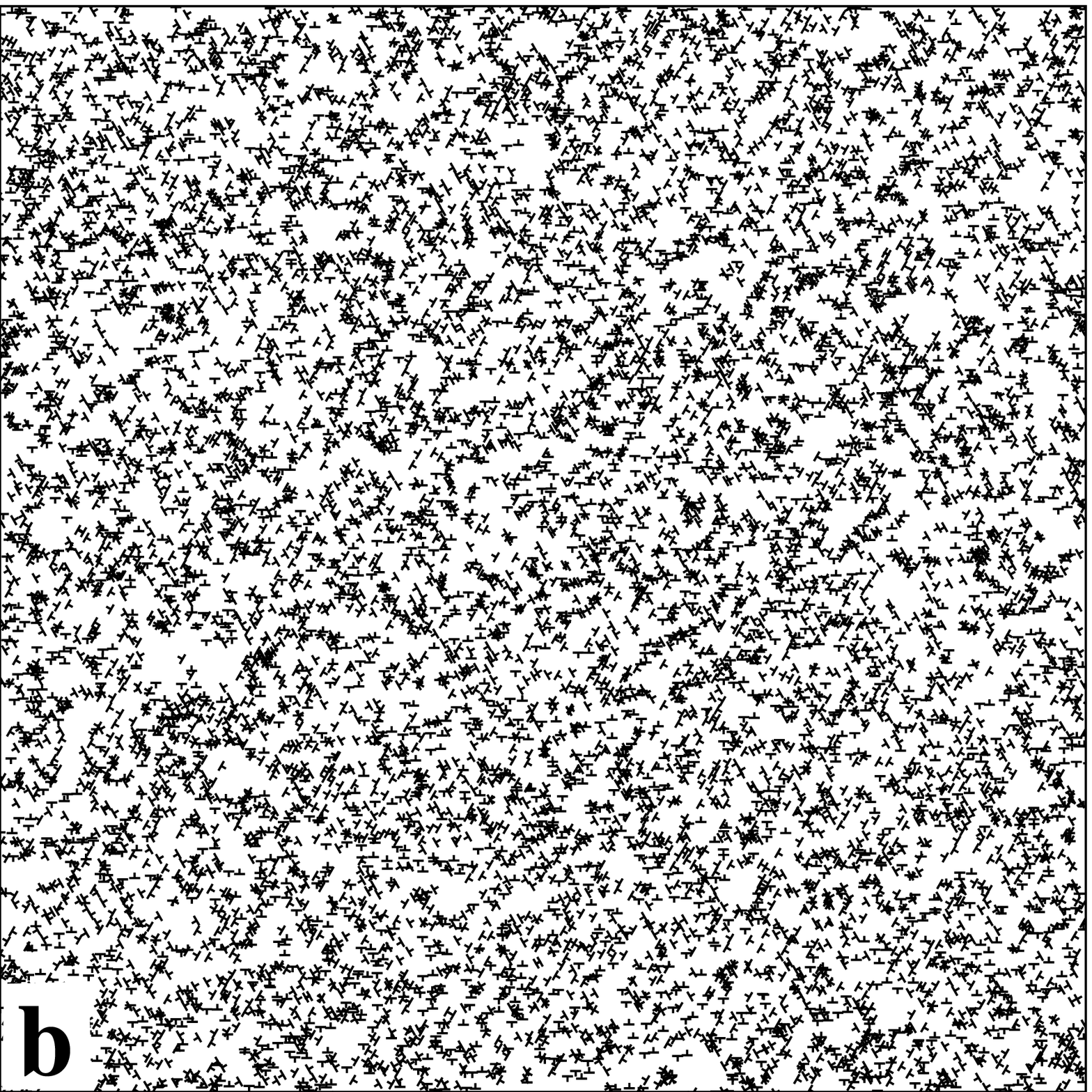}\\
\includegraphics[width=0.2\textwidth]{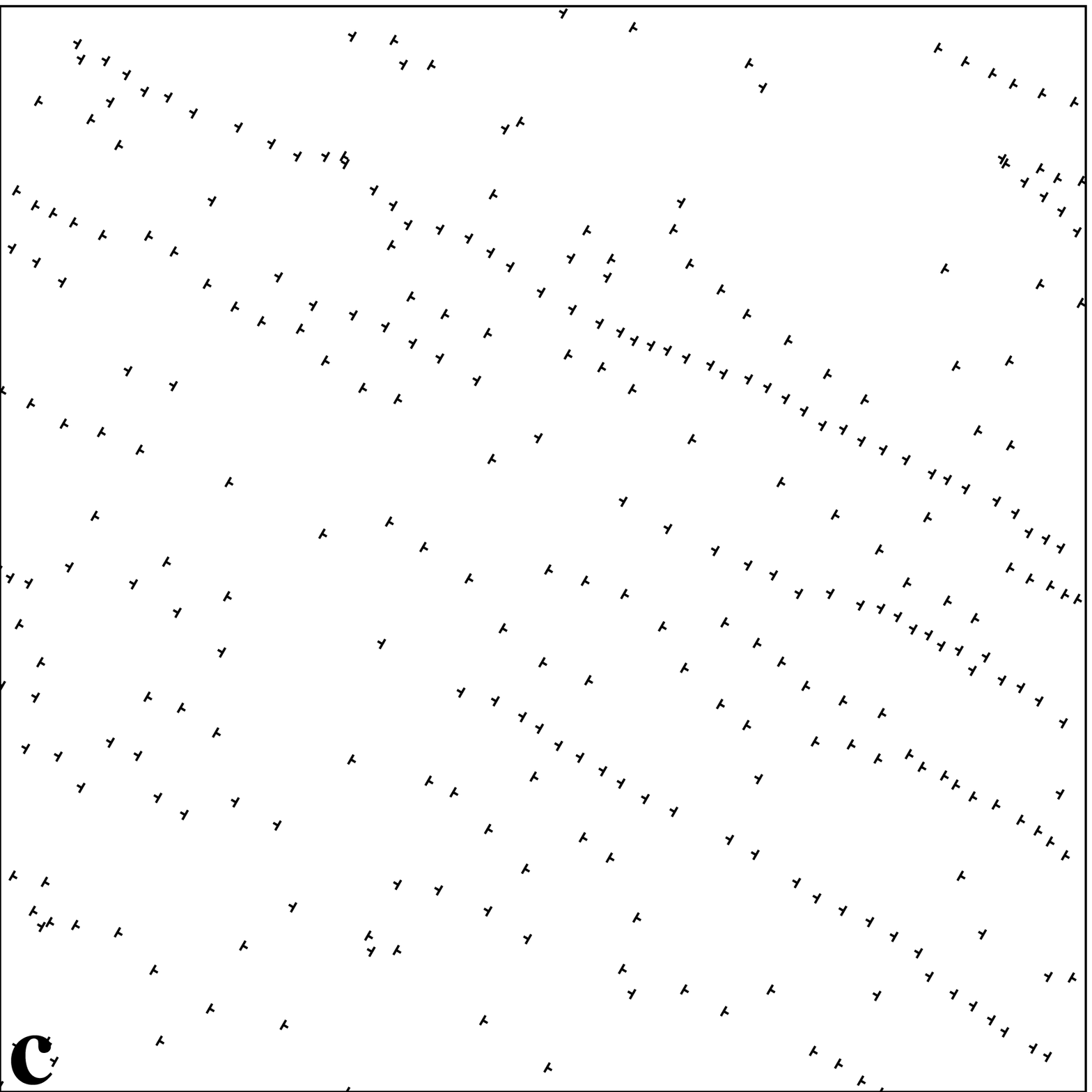}
\includegraphics[width=0.2\textwidth]{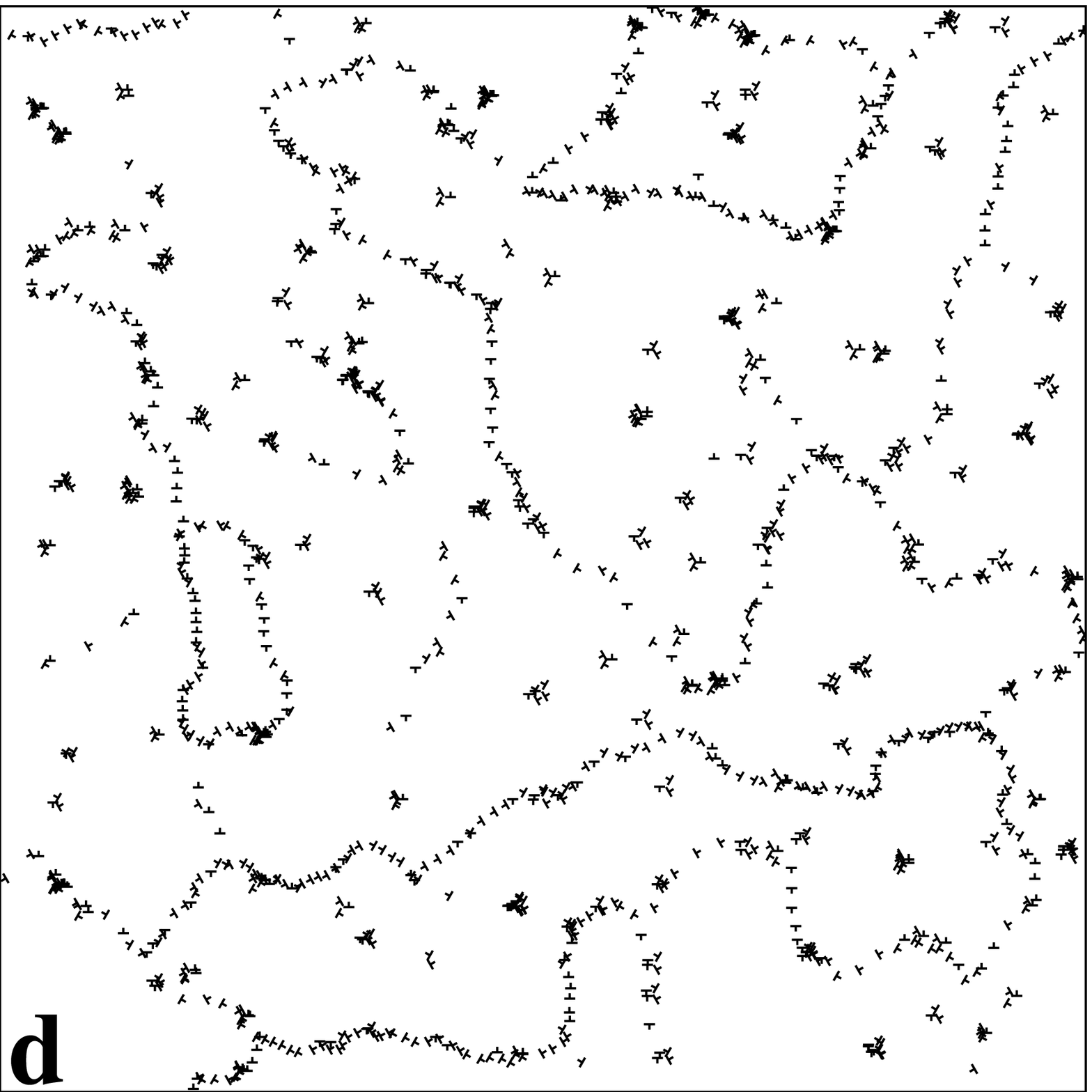}
\end{center}
\caption{Snapshots of relaxed dislocation configurations without
climb
($\Gamma_c=0$): (a) single slip (b) 3-slip,  and typical wall structures with
climb ($\Gamma_c/\Gamma_b=0.1$): (c) single slip, (d) 3-slip.}
\label{fig:confs}
\end{figure}

In order to study the morphology of dislocations, we considered
dynamics without and with climb, and in either case we took 1 as well
as 3 glide axes.  Simulation started out from random initial
conditions of 10000 dislocations, specifically, axes were placed
independently, with a uniform distribution, and the sign of the
Burgers vector was also randomly chosen such that for each axis
direction they compensated each other.  No shear is imposed through
external traction, as we intend to study dislocations originating from
some internal inhomogeneities. If climb is present, dislocations can
get arbitrarily close to each other, so we annihilated dislocations of
opposite Burgers vectors in the same box.  Figures \ref{fig:confs}a-d
show typical configurations, without climb (a,b) and with climb and
annihilation (c,d), in each case for 1 and 3 slip axes, respectively.
Configurations (a,b) are near equilibrium, they show diffuse
distributions, but in the single slip case the formation of walls is
apparent, in accordance with recent analytic results \cite{groma06}.
On (c,d) transient states are snapshot, where the predominance of
walls is apparent.  Dislocations eventually annihilate, in agreement
with the classic prediction \cite{halperin78} of dipolized equilibrium
well below the melting point.  As we have reported in \cite{bako06},
walls form cells in the 3-slip case as seen in Fig.\ \ref{fig:confs}d.
The picture shows a close resemblance to some dislocation patterns in
2d dusty plasmas \cite{quinn01}, highlighting the experimental
relevance of our simulation.  Furthermore, our results are in
qualitative concordance with recent experimental studies
\cite{rudolphetal05}, where, in off stoichiometric GaAs, cell
formation was attributed to increased climb mobility.  So we can
conclude that first and foremost the presence of climb, but also the
number of slip axes, strongly affect the morphology of dislocations.
It is common for all cases presented here, that random initial
conditions result in disordered configurations, calling for studies
into glassy properties of dislocation systems.

Aiming at quantifying the glassy character of dislocation dynamics we focus on
ageing (see \cite{biroli05}).  In particular, we study the  dependence on the
waiting time of the correlation, or overlap, function
\begin{eqnarray}
C(t_{\text w}, t) = \frac{1}{N_{\text{d}}}\sum_{i = 1}^{N_{\text{d}}}\exp\left [
-|{\bf r}_i(t + t_{\text{w}}) - {\bf r}_i(t_{\text{w}})|/r_0
\right ],
\label{eq:Cfunc}
\end{eqnarray}
where somewhat arbitrarily $r_0=5/\sqrt{N_{\text{d}}}$, and the effective
diffusion coefficient
\begin{eqnarray}
D(t_{\text{w}},t) = \frac{1}{N_{\text{d}}t}\sum_{i = 1}^{N_{\text{d}}}|{\bf
r}_i(t + t_{\text{w}}) -
{\bf r}_i(t_{\text{w}})|^2,
\label{eq:Diff}
\end{eqnarray}
where $N_{\text{d}}$ is the number of dislocations remaining at the final
observation time.  Here $C$ and $D$  should be conceived as time $t$ dependent
quantities, recorded after a waiting time $t_{\text{w}}$.

\begin{figure} 
\begin{center} 
\includegraphics[height=0.35\textwidth,width=0.2\textwidth,angle=-90]{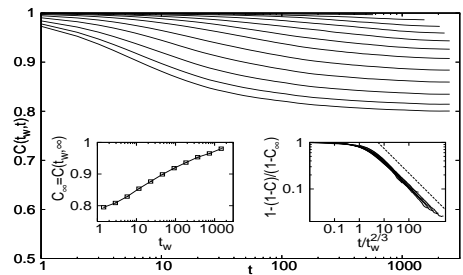} 
\end{center}
\caption{Correlation function $C(t_{\text w},t)$ for 3-slip,
without climb at $T=0$ as function of $t$ for different $t_w=1,2,...,2048$.
Increasing $t_w$
leads to slower decay. Left and right inset show the asymptote and the scaling
 form, resp.}
\label{fig:C-nc} 
\end{figure}

\begin{figure} 
\begin{center} 
\includegraphics[height=0.342\textwidth,angle=-90]{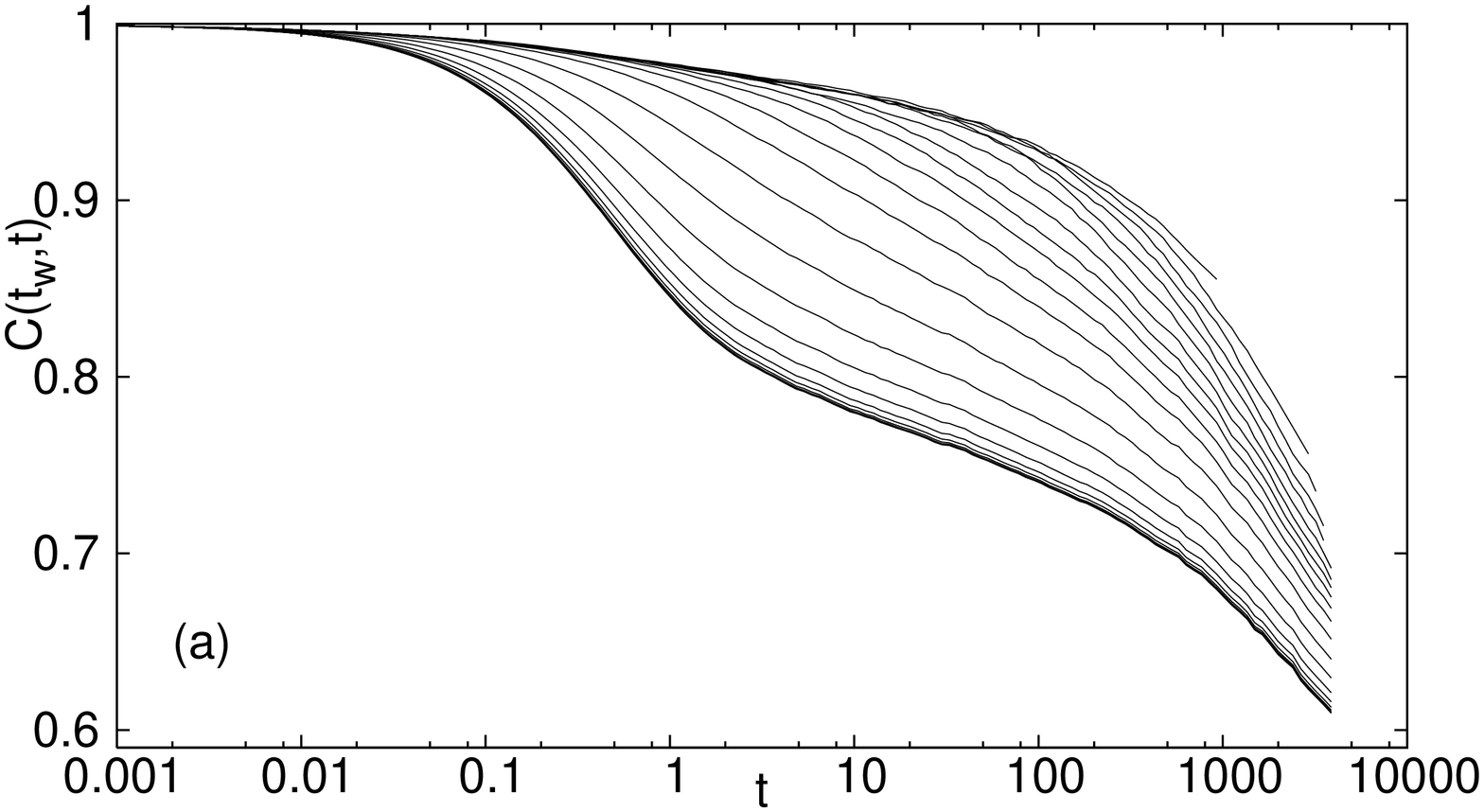}
\includegraphics[height=0.35\textwidth,angle=-90]{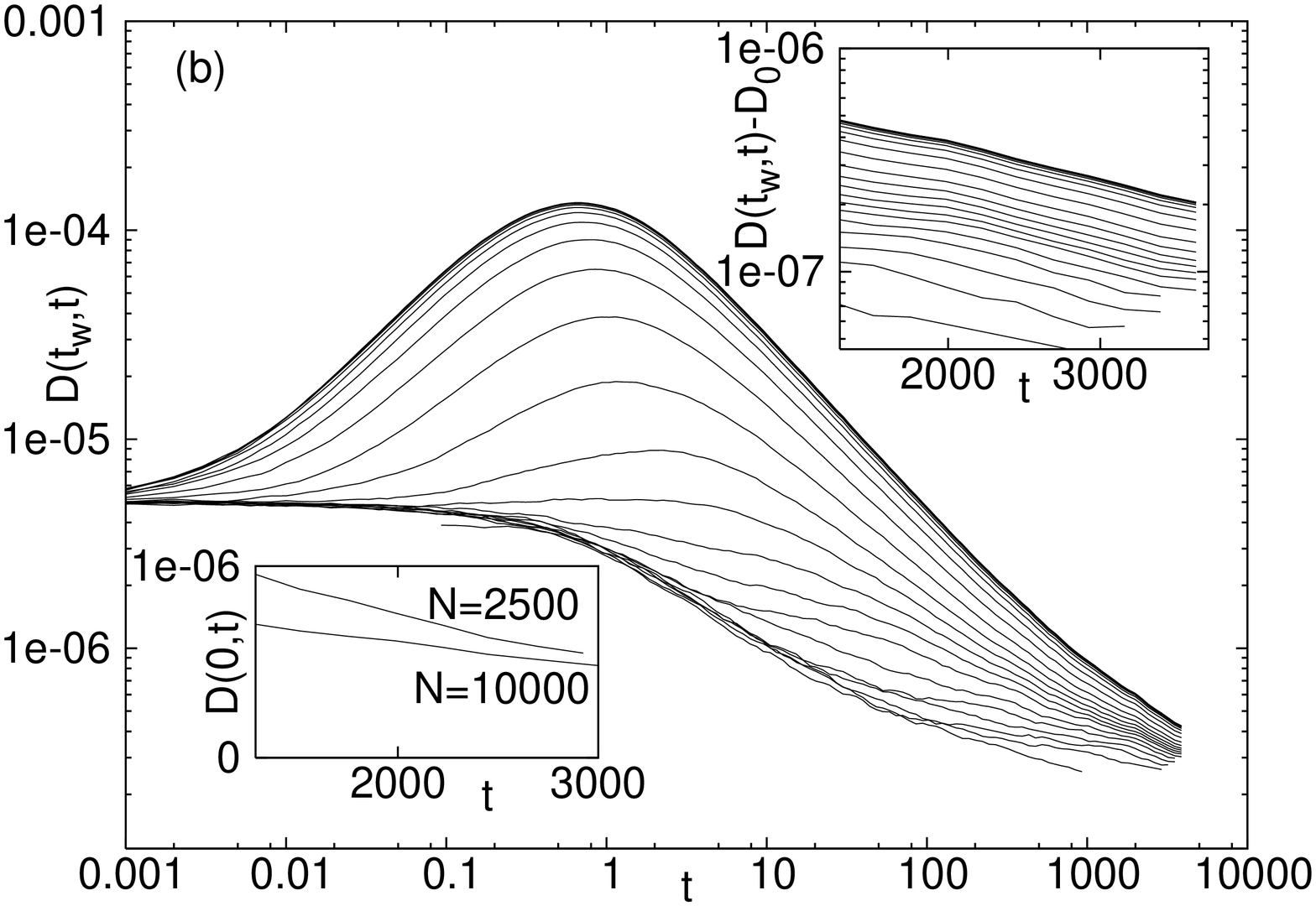}
\end{center}
\caption{ Correlation function $C(t_{\text w},t)$  (a) and diffusion coefficient
$D(t_{\text w},t)$  (b) for 3-slip, without climb, $T=0.025$, as function of $t$
for various  $t_{\text w}$-s. Upper inset in (b) shows the decay to the
asymptotic $D_0=2.2\, 10^{-7}$ and lower inset gives $D(0,t)$ for two system
sizes.}
\label{fig:CD-nc} 
\end{figure}

First we consider the situation without climb, with 3 slip axes, at
zero temperature, and plot the correlation function on Fig.\
\ref{fig:C-nc}.  The asymptotic value $C_\infty(t_{\text
  w})=C(t_{\text w},\infty)$ markedly depends on $t_{\text{w}}$ (left
inset).  Generically, beyond the inflexion point correlations decay
exponentially, followed by a power form $t^{-\beta}$.  Thus it is
natural to consider the curves $1-(1-C)/(1-C_\infty)$ decaying in time
$t$ from $1$ to $0$, which are found to collapse approximately to a
master curve as function of $t/t_{\text{w}}^{2/3}$.  This is shown on
the right inset, where the straight line indicates power law with
$\beta\approx 0.54$.  Note that the fact that $t_w$ and $t$ have
different exponents contradicts standard scaling formulas in ageing
(see \cite{biroli05}), an anomaly for which we do not have an
explanation and requires further studies.  It is worth recalling here
that power decay of the elastic energy was found before in
\cite{csikor05}.  So we can conclude here that there is a strong
analogy with spin glasses inasmuch that there is a quenched randomness
in the placement of the slip axes, which confine dislocations along
the entire dynamics, and the relaxation towards a disordered ground
state exhibits ageing.
Next we study the dynamics of the 3 slip system without climb in the
presence of a small Langevin noise, $T=0.025$.  Then the system never
fully relaxes, but before some stationary state is reached it also
exhibits ageing, as demonstrated in Fig.\ \ref{fig:CD-nc}.  Plot (a)
shows the correlation function decaying expectedly to zero (we did not
reach the asymptote due to limitations in computational power) for a
sequence of waiting times.  On (b) the diffusion coefficient is shown,
where lower lying curves correspond to increasing $t_{\text{w}}$.
Asymptotically we found $D(t_{\text{w}},t)\approx D_0 + D_1(t_{\text
  w}) t^{-\gamma}$ with $D_0\approx 2.2\, 10^{-7}, \gamma\approx 0.8$,
see upper right log-log inset.  The power decay to $D_0$ can be
considered as a manifestation of slowed-down dynamics, which may
appear for intermediate $t$-s as effectively subdiffusive.  To test
whether the remanent diffusion constant $D_0$ is physically relevant
or only due to finite size effects, we halved the effective linear
system size by running a simulation with $N_{\text d}=2500$.  The
lower left inset shows that the asymptotic $D_0$ does not change
significantly for varying system size, so ordinary diffusion indeed
prevails. 
 
\begin{figure}
\begin{center}
\includegraphics[height=0.22\textwidth,angle=-90]{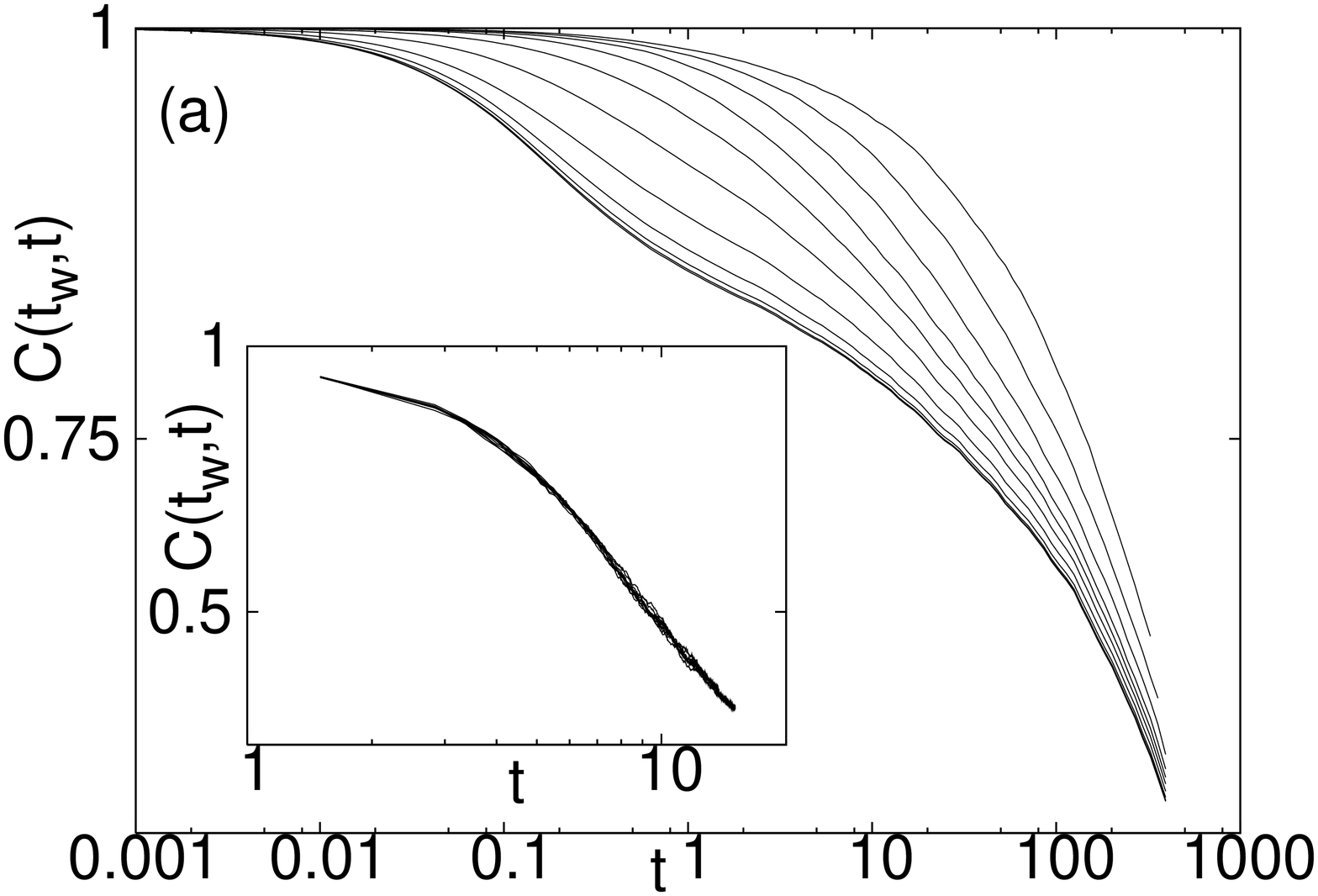} \
\includegraphics[height=0.22\textwidth,angle=-90]{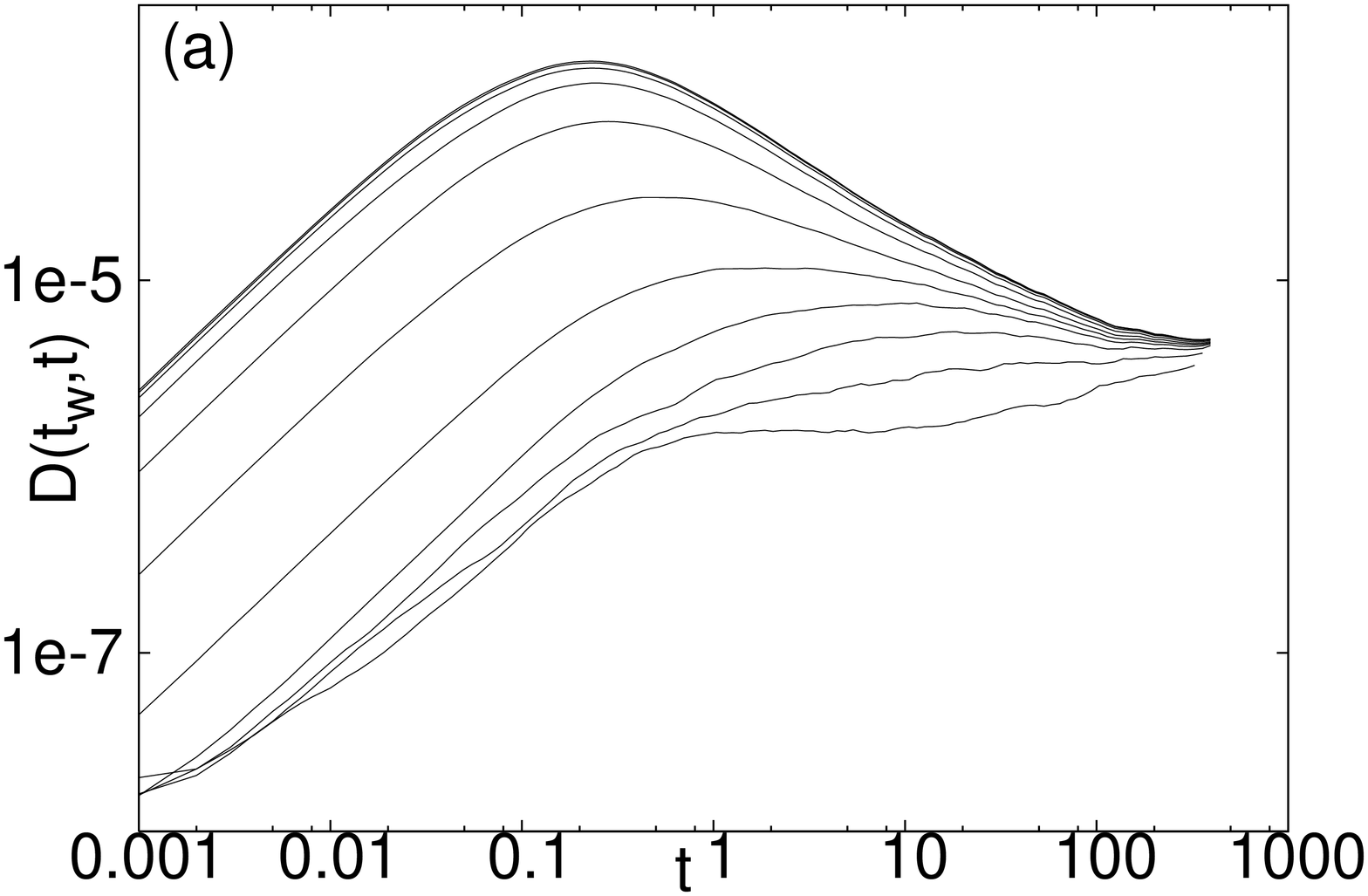}
\end{center}
\caption{ Correlation function (a) and diffusion coefficient (b) for 3-slip,
with climb ($\Gamma_c/\Gamma_g=0.1$) and annihilation, otherwise like in Fig.\
\ref{fig:CD-nc}. Inset in (a) is the plot of the correlation function  at a
sufficiently high temperature
showing that the dependence on the waiting time $t_{\text w}$ is practically
lost. } \label{fig:CD-c}
\end{figure}

Finally, we considered dislocations with climb and annihilation for 3
slip axes.  Here the number of dislocations $N_{\text d}$ is a
decreasing function of time, and after an initial transient the
dislocations order into cellular structures, whose characteristic
length scale grows while dislocations annihilate.  Figure
\ref{fig:CD-c} shows the ageing effect in the overlap function and the
diffusion coefficient, different curves corresponding to different
values of the waiting time. A small thermal force was included,
presumably having no significant effect on dislocations in cell walls.
The inset shows the correlation function at a sufficiently high
temperature that suppressed ageing.
As we neglected the creation of dislocations, the process we followed
is essentially a transient before all dislocations annihilate.
Furthermore, due to climb the dynamics is not constrained to the
randomly placed glide axes, so disorder in the morphology is
essentially due to random initial conditions.  This situation is
reminiscent to the transient observed in systems with non-disordered
equilibrium states but showing a transient with irregular domain
coarsening if quenched from a sufficiently high temperature. While
systems with interrupted ageing are usually not considered as glasses,
their transients do show ageing and, on short time scales, further
glassy characteristics were detected recently \cite{krzakala05}.  A
common property of those systems is the asymptotic scaling of the
characteristic domain size $\ell$ with time as $\ell\propto t^{1/z}$.
For instance, $z=2,3$ corresponds to the 3d Ising model with Glauber
and Kawasaki dynamics, and the logarithmic growth in the random field
model can be associated with $z=\infty$, see \cite{biroli05}.  Earlier
we found numerically that the number of dislocations decreased
approximately like $N_{\text d}\propto t^{-\delta}$ with $\delta$
approximately $1/3$ \cite{bako06}.  Hence, based on the scaling of all
lengths by the dislocation number we can conclude that the dynamical
exponent here is in the order of $z\approx 6(1)$, where the error
estimate is subjective.  This relatively large exponent means that the
glassy transient lasts for comparatively long times and may explain
why the cellular structure can persist under experimental conditions.

We can summarize the physical picture of disordered dislocation
systems as follows.  In the absence of climb, where the external
disorder is in the random but fixed placement of glide axes, there is
an analogy with spin glasses in that the dislocation system has a
multitude of near ground states, to one of which it relaxes.  No
increasing length scale was discerned at this stage of simulation.  On
the other hand, with climb, the ever growing cellular structure
resembles the domain growth in, say, the Ising model launched after a
quench from a paramagnetic state and showing glass-like dynamics. In
each case there is a pronounced ageing effect in both the correlation
function and the diffusion coefficient at zero and small temperatures.
Interestingly, we did not encounter dislocation states analogous to
structural glasses, probably because dislocations moving without
constraints can annihilate.  Our results are, however, strongly
suggestive that we can speak about dislocation glasses, where the
types of glasses are distinguished by different morphologies.

Financial support by the European Community's Human Potential
Programme under Contract No.\ MRTN-CT-2003-504634 [SizeDepEn] and the
Hungarian Scientific Research Fund (OTKA) under Contract No.\ T~043519
and TS~044839 are gratefully acknowledged.

\end{document}